\documentclass[prc,twocolumn,showpacs,amsmath,amssymb]{revtex4}
\usepackage[toc,page,header]{appendix}
\usepackage{times}
\usepackage{graphicx}
\usepackage{dcolumn}
\usepackage{bm}
\usepackage{amstext}

\begin{document}

\title{Application of relativistic mean field and effective field theory 
densities to scattering observables for Ca isotopes}

\author{M. Bhuyan$^{1,2}$, R. N. Panda$^3$, T. R. Routray$^{2}$  and 
S. K. Patra$^1$
}

\affiliation{$^1$ Institute of Physics, Sachivalaya Marg, Bhubaneswar-751 005, 
India. \\ 
$^2$ School of Physics, Sambalpur University, Jyotivihar, Burla-768 019, 
India. \\
$^3$ Dept. of Physics, ITER, Siksha O Anusandhan University, 
Bhubaneswar-751 030, India.
}

\date{\today}

\begin{abstract}
In the frame work of relativistic mean field (RMF) theory, we have calculated 
the density distribution of protons and neutrons for $^{40,42, 44,48}Ca$ 
with NL3 and G2 parameter sets. The microscopic proton-nucleus optical 
potentials for $p+^{40,42,44,48}Ca$ systems are evaluated from the Dirac 
NN-scattering amplitude and the density of the target nucleus using 
Relativistic-Love-Franey and McNeil-Ray-Wallace parametrizations. We 
have estimated the scattering observables, such as elastic differential scattering 
cross-section, analyzing power and the spin observables with the relativistic 
impulse approximation (RIA). The results have been compared with the experimental data for 
few selective cases and find that the use of density as well as the 
scattering matrix parametrizations are crucial for the theoretical prediction.
\end{abstract}


\maketitle

\section{Introduction}

Study of the nuclear reactions is a challenging subject
of nuclear physics both in theory and laboratory. This is useful to explain 
the nuclear structure of stable as well as exotic nuclei. The Nucleon- Nucleus
interaction provides a wide source of information to determine the nuclear 
structure including spin, isospin, momenta, densities and gives a clear path 
towards the formation of exotic nuclei in the laboratory. In this context 
the study of elastic scattering of Nucleon-Nucleus is more interesting 
than that of Nucleus-Nucleus at different energies. One of the theoretical 
method to study such type of reactions is the "Relativistic 
Impulse Approximation" (RIA). It is a microscopic theory where the Dirac 
optical potential is constructed from the Lorentz invariant Nucleon-Nucleon
(NN) amplitudes 
obtained from relativistic meson exchange models. The basic ingredients 
in this approach are the NN scattering amplitude and the nuclear scalar 
and vector densities \cite{zpli08} of the target nucleus. This approach 
can be extended to elastic scattering of composite particles
\cite{amorim92}. In this context proton-Nucleus (p-A) elastic scattering 
is of particular interest because of its relative simplicity with which it 
provides a satisfactory description of the reaction dynamics.

One useful application of RIA is to generate microscopic optical
potential to study the elastic and inelastic scattering of nucleons
for unstable proton- and/or neutron- rich nuclei. The RIA folding procedure can
also be extended to calculate microscopic optical potentials for exotic
nuclei using relativistic mean field formalism \cite{tod03, meng06}.

The first theoretical introduction to elastic scattering  
was given by Chew \cite{chew01} almost six decades ago. For a wide range 
of energy interval, Impulse Approximation (IA) produces the main 
qualitative description on quasi-elastic scattering for $A\leq 64$ nuclei 
\cite{bala68}. During the same time, Glauber \cite{glau55} studied the 
reaction dynamics of the composite system at low energies but this model 
is unable to predict extension of quasi-elastic scattering. Further the 
generalized Glauber formula and the unitarized impulse approximation 
\cite{fedd63,mahu78} were circumvented. But the development of RIA opens 
a path to study the above mentioned scattering phenomenon for both the elastic and 
the quasi-elastic particles. 
This field of research was further strengthened by the experimental 
evidences of cross-section and analyzing power for the scattering systems 
$\overrightarrow {p}+^{12}C$, ${p}+^{9}B$ and ${p}+^{16}O$ 
at 200 MeV, which were measured over a wide range of momentum transfer 
$ > 6fm^{-1}$ at IUCF \cite{meye81,bach80}. Recent study of proton 
nucleus elastic scattering within modified (Coulomb) Glauber model 
\cite{khan07,khan09} and global Dirac optical potential \cite{hama90} 
have motivated us to study the elastic scattering phenomenon. For convenience, we 
consider Ca isotopes as targets and $p$ as a projectile, because Ca satisfies 
the relativistic mean field nuclear structure model accurately without recoil 
correction to the Dirac scattering equation. 

In the present paper, our aim is to calculate the nucleon-nucleus 
elastic differential scattering cross-section ($\frac{d\sigma}{d\Omega}$) 
and other related physical quantities, like optical potential ($U_{opt}$), analyzing power 
($A_y$) and spin rotation parameter ($Q-$value) using relativistic mean 
field (RMF) and recently proposed effective field theory motivated 
relativistic mean field (E-RMF) densities. These are 
obtained from the successful NL3 \cite{lala97} and the advanced G2 
\cite{tang96} parameter sets, which are given in Section II. 
In the Sections III and IV, the details of target 
densities folded with the NN-amplitude for various energetic proton 
projectile with Relativistic-Love-Franey (RLF) \cite{witz85,mur87} and McNeil-Ray-Wallace 
(MRW) parametrizations \cite{neil83} for $^{40,42,44,48}Ca$ are given. 
In these sections we have outlined the 
expressions for the differential elastic scattering cross-section, analyzing power 
and spin observables. Section V describes the results obtained from our 
calculations. Finally a brief summary and conclusions are given in 
the Section VI for the present work.

\section{The RMF and E-RMF formalisms}

A documentation of RMF  and E-RMF  formalisms are available in Refs.  
\cite{patra91} and \cite{tang96,patra01a} respectively for both finite and 
infinite nuclear matter. Here only the energy density functional and associated
expressions for the densities are presented\cite{ser97,fur96}.  

\begin{widetext}
\begin{eqnarray}
\mathcal{E}(\mathbf{r}) & = & \sum_\alpha \varphi_\alpha^\dagger 
\Bigg\{ -i \mbox{\boldmath$\alpha$} \!\cdot\! \mbox{\boldmath$\nabla$} 
+\beta (M -\Phi) + W +\frac{1}{2}\tau_{3}R+\frac{1+\tau_3}{2} A -\frac{i}{2M} 
\beta \mbox{\boldmath$\alpha$}\!\cdot\! \left( f_v \mbox{\boldmath$\nabla$} W 
+ \frac{1}{2}f_\rho\tau_3 \mbox{\boldmath$\nabla$}R+\lambda 
\mbox{\boldmath$\nabla$}A\right) \nonumber\\
&& +\frac{1}{2M^2}\left(\beta_s + \beta_v \tau_3 \right)
\Delta A \Bigg\} \varphi_\alpha \null + \left(\frac{1}{2} 
+\frac{\kappa_3}{3!}\frac{\Phi}{M}+\frac{\kappa_4}{4!}\frac{\Phi^2}{M^2}\right )
\frac{m_{s}^2}{g_{s}^2}\Phi^2-\frac{\zeta_0}{4!} \frac{1}{ g_{v}^2 }W^4\null
+\frac{1}{2g_{s}^2}\left( 1 + \alpha_1\frac{\Phi}{M}\right)
\left(\mbox{\boldmath $\nabla$}\Phi\right)^2 \nonumber\\
&& -\frac{1}{2g_{v}^2}\left(1+\alpha_2\frac{\Phi}{M}\right) 
\left(\mbox{\boldmath $\nabla$} W\right)^2 \null
-\frac{1}{2}\left(1 + \eta_1\frac{\Phi}{M} 
+\frac{\eta_2}{2}\frac{\Phi^2 }{M^2} \right)\frac{{m_{v}}^2}{{g_{v}}^2} W^2
-\frac{1}{2g_\rho^2} \left( \mbox{\boldmath $\nabla$}R\right)^2 
-\frac{1}{2} \left(1 + \eta_\rho \frac{\Phi}{M} \right) \frac{m_\rho^2}{g_\rho^2} 
R^2 \null \nonumber\\
&& -\frac{1}{2e^2}\left(\mbox{\boldmath $\nabla$} A\right)^2 
+\frac{1}{3g_\gamma g_{v}}A \Delta W + \frac{1}{g_\gamma g_\rho}A \Delta R ,
\end{eqnarray}
\end{widetext}

\noindent
where the index $\alpha$ runs over all occupied states $\varphi_\alpha (%
\mathbf{r})$ of the positive energy spectrum, $\Phi \equiv g_{s} \phi_0(%
\mathbf{r})$, $W \equiv g_{v} V_0(\mathbf{r})$, $R \equiv g_{\rho}b_0(%
\mathbf{r})$ and $A \equiv e A_0(\mathbf{r})$.

The terms with $g_\gamma$, $\lambda$, $\beta_{s}$ and $\beta_{v}$ take care 
of the effects related with the electromagnetic structure of the pion and 
the nucleon (see Ref. \cite{fur96}). Specifically, the constant $g_\gamma$ 
concerns the coupling of the photon to the pions and the nucleons through 
the exchange of neutral vector mesons. The experimental value is $%
g_\gamma^2/4\pi = 2.0$. The constant $\lambda$ is needed to reproduce the 
magnetic moments of the nucleons, defined by
\begin{eqnarray}
\lambda = \frac{1}{2} \lambda_{p} (1 + \tau_3) + \frac{1}{2} \lambda_{n} (1
- \tau_3) ,  \label{eqFN2}
\end{eqnarray}
with $\lambda_{p}=1.793$ and $\lambda_{n}=-1.913$, the anomalous magnetic 
moments of the proton and the neutron, respectively. The terms with 
$\beta_{s}$ and $\beta_{v}$ contribute to the charge radii of the nucleon 
\cite{fur96}.

The energy density contains tensor couplings, scalar-vector and 
vector-vector meson interactions in addition to the standard scalar 
self-interactions $\kappa_{3}$ and $\kappa_{4}$. Thus, the E-RMF formalism can 
be interpreted as a covariant formulation of density functional theory 
as it contains all the higher order terms in the Lagrangian, obtained by 
expanding it in powers of the meson fields. The terms in the Lagrangian 
are kept finite by adjusting the parameters. Further insight into the 
concepts of the E-RMF model can be obtained from Ref. \cite{fur96}. It may 
be noted that the standard RMF Lagrangian is obtained from that of the 
E-RMF by ignoring the vector-vector and scalar-vector cross interactions, 
and hence does not need a separate discussion.

In each of the two formalisms (E-RMF and RMF), the set of coupled equations 
are solved numerically by a self-consistent iteration method. The baryon, 
scalar, isovector, proton and tensor densities are

\begin{eqnarray}
\rho(r) &=&
\sum_\alpha \varphi_\alpha^\dagger(r) \varphi_\alpha(r) \,,
\label{eqFN6} \\[3mm]
\rho_s(r) &=&
\sum_\alpha \varphi_\alpha^\dagger(r) \beta \varphi_\alpha(r) \,,
\label{eqFN7} \\[3mm]
\rho_3 (r) &=&
\sum_\alpha \varphi_\alpha^\dagger(r) \tau_3 \varphi_\alpha(r) \,,
\label{eqFN8} \\[3mm]
\rho_{\rm p}(r) &=&
\sum_\alpha \varphi_\alpha^\dagger(r) \left (\frac{1 +\tau_3}{2}
\right)  \varphi_\alpha(r) \,,
\label{eqFN9}  \\[3mm]
\rho_{\rm T}(r) &=&
\sum_\alpha \frac{i}{M} \mbox{\boldmath$\nabla$} \!\cdot\!
\left[ \varphi_\alpha^\dagger(r) \beta \mbox{\boldmath$\alpha$}
\varphi_\alpha(r) \right] \,,
\label{eqFN10} \\[3mm]
\rho_{\rm T,3}(r) &=&
\sum_\alpha \frac{i}{M} \mbox{\boldmath$\nabla$} \!\cdot\!
\left[ \varphi_\alpha^\dagger(r) \beta \mbox{\boldmath$\alpha$}
\tau_3      \varphi_\alpha(r) \right] \,.
\end{eqnarray}

These densities are obtained from the RMF and E-RMF 
formalisms with NL3 \cite{lala97} and G2 \cite{tang96} parametrizations.
We refer the readers to Refs.  \cite{patra91,patra01a} for numerical 
details and ground state equations for finite nuclei.

\section{The Nucleon-Nucleon scattering Amplitude}

The non-linear relativistic impulse approximation (RIA) involves mainly two 
steps \cite{lace83,hama83,mer83} of calculation. Basically a particular 
set of Lorentz covariant function \cite{neil83}, which multiply with the 
so called Fermi invariant Dirac matrix represent the Nucleon-Nucleon 
NN-scattering amplitudes. This functions are then folded with the target 
densities of proton and neutron from the relativistic 
Langragian for NL3 and G2 parameter sets to produce a first order 
complex optical potential. The invariant NN-scattering operator 
${\cal F}$ can be written in terms of five complex functions (the five 
terms involves in the proton-proton pp and neutron-neutron nn scattering). 
In general RIA, the function ${\cal F}$ can be expressed as 
\cite{lace83,hama83,mer83},

\begin{eqnarray}
{\cal F}({\it q}, {\it E}) =\sum_{L=S}^{PS} {\cal F}^{L} ({\it q}, {\it E}) 
\lambda^{L}_{(0)}.\lambda^{L}_{(1)},
\end{eqnarray}
where $\lambda^L$ stands for Dirac operator and (0) and (1) for the incident and 
struck nucleons respectively. The 
S, V, T, A and PS  stands for scalar, vector, axial vector, tensor 
and pseudoscalar.
The dot product (.) implies all Lorentz indices are contracted. The Dirac 
spinor is defined the initial and final two nucleons by taking the matrix 
elements of ${\cal F}$, which represent the NN-scattering amplitudes. The 
function ${\cal F}^{L}$ are determined by equating the resultant amplitude 
(in center of mass frame) to the empirical amplitude, which is conventionally 
expressed in term of the non-relativistic Wolfenstein amplitudes 
$A_{1}, A_{2},....A_{5}$ \cite{neil83}. Since there are  five complex 
invariant amplitudes and  $A_{1}, A_{2},....A_{5}$ are five Wolfenstein 
amplitudes, the relation among them is determined by a $5\times5$ non-singular 
matrix, whose inversion is straight forward. However ${\cal F}$ is an 
operator in the two particle Dirac space and the component cancelled out due 
to isospin and parity invariance and we get only 44 components 
\cite{tjon85}. From the above, it is clear, to specify that ${\cal F}$ is 
not unique. In other words, there are infinite number of operators ${\cal F}$ 
with same five on-shell but different negative (energy) elements. The 
expression of ${\cal F}$ cannot predict reasonanle result at lower energy region. 
To avoid the limitation, the pseudoscalar ${\cal F}^{PS}$ is replaced by the 
pseudovector invariant, and is expressed as,

\begin{eqnarray}
{\cal F}^{PS}\gamma^{5}_{(0)}\gamma^{5}_{(1)}=-{\cal F}^{PV}\frac{\gamma^{5}_{(0)}}{2M}
\frac{\gamma^{5}_{(1)}}{2M}.
\end{eqnarray}

The meson-nucleon couplings are complex, with a real part ${\it g}^{2}_{i}$ 
and an imaginary part $\overline{{\it g}^{2}_{i}}$, which can be decomposed 
into two parts,

\begin{widetext}
\begin{eqnarray}
<k^{'}_{0}k^{'}_{1} \mid {\cal F} \mid k_{0}k_{1}> &=& <k^{'}_{0}k^{'}_{1}
\mid {\it t}({\it E})\mid k_{0}k_{1}>
+(-1)^{T}<k^{'}_{0}k^{'}_{1}\mid {\it t}({\it E})\mid k_{0}k_{1}>,
\end{eqnarray}
\end{widetext}
where {\it t}({\it E}) is the lowest order meson and T is the total isospin 
of the two nucleon state. The calculation of the one-meson-exchange from 
Feynman diagram \cite{witz85} is represented as,

\begin{eqnarray}
{\it g}_{i}(\frac{\Lambda^{2}_{i}}{{\it q}^{2}+{\Lambda^{2}_{i}}})\lambda^{{\it L}_{(i)}}(\tau)^{I_{i}}, 
\end{eqnarray}
with ${\it L}_{(i)}$ denotes spin and parity of the $i^{th}$ meson and 
${I_{i}}$ = (0, 1) is the meson's isospin. Here we neglect the energy transfer 
${\it q}^{0}$ carried by the meson for different masses and cut off 
parameters in the real and imaginary parts of the amplitude in Eqn. (9). 
The contribution of $i^{th}$-meson to the NN-scattering amplitude 
by taking all kinematic is given as,

\begin{widetext}
\begin{eqnarray}
\overline{U_{0^{'}}} . \overline{U_{1^{'}}}{\cal F}_{i}U_{0}U_{1}
&\propto& \frac{{\it g}^{2}_{i}}{q^{2}+m^{2}_{i}}{\left(\frac{\Lambda^{2}_{i}}{{q}^{2}
+{\Lambda^{2}_{i}}}\right)}^{2}
{\lbrace\tau_{0}.\tau_{1}}\rbrace^{I_{i}}\overline{U_{0^{'}}}\lambda^{L_{i}}{U_{0}}.
\overline{U_{1^{'}}}\lambda^{L_{i}}{U_{1}}\nonumber\\
&& +(-1)^{T}\frac{{\it g}^{2}_{i}}{Q^{2}+m^{2}_{i}}
{\left(\frac{\Lambda^{2}_{i}}{{Q}^{2}+{\Lambda^{2}_{i}}}\right)}^{2}
{\lbrace\tau_{0}.\tau_{1}}\rbrace^{I_{i}}\overline{U_{1^{'}}}\lambda^{L_{i}}{U_{0}}.
\overline{U_{0^{'}}}\lambda^{L_{i}}{U_{1}},
\end{eqnarray}
Here the direct and exchange momentum transfer are $q=k_{0}^{'}-k_{0}$ 
and $Q=k_{1}^{'}-k_{1}$. The first term in Eq.(13), which is already of the 
form of Eq.(9), can easily identify the contribution of ${\cal F}^{L}$. The 
second term is unlike to this form, so we rewrite this as,

\begin{eqnarray}
\overline{U_{0^{'}}} . \overline{U_{1^{'}}}{\cal F}_{i}U_{0}U_{1}
&\propto& \frac{{\it g}^{2}_{i}}{q^{2}+m^{2}{_i}}{\left(\frac{\Lambda^{2}_{i}}{{q}^{2}
+{\Lambda^{2}_{i}}}\right)}^{2}{\lbrace\tau_{0}.\tau_{1}}\rbrace^{I_{i}}
\overline{U_{0^{'}}}\lambda^{L_{i}}{U_{0}}.
\overline{U_{1^{'}}}\lambda^{L_{i}}{U_{1}}\nonumber\\
&& +(-1)^{T}\sum_{L'} {\it B}_{L_{i}}\frac{{\it g}^{2}_{i}}{Q^{2}+m^{2}_{i}}
{\left(\frac{\Lambda^{2}_{i}}{{Q}^{2}+{\Lambda^{2}_{i}}}\right)}^{2}
{\lbrace\tau_{0}.\tau_{1}}\rbrace^{I_{i}}\overline{U_{0^{'}}}\lambda^{L'_{i}}{U_{0}}.
\overline{U_{1^{'}}}\lambda^{L'_{i}}{U_{1}},
\end{eqnarray}
\end{widetext}
where the transformation matrix is given as,

\begin{eqnarray}
{\it B}_{L,L{'}} = \frac{1}{8}
\left[ \begin{array}
{ccccc} 
2 & 2 & 1 & -2 & 2 \\ 
8 & -4 & 0 & -4 & -8 \\ 
24 & 0 & -4 & 0 & 24 \\ 
-8 & -4 & 0 & -4 & 8 \\
2 & -2 & 1 & 2 & 2 
\end{array} \right].
\end{eqnarray}

The row and columns are labeled in the order of {\it S, V, T, A, PS}. 
The contribution to the Lorentz invariants 
(${\cal F}_{L}$) in simpler forms are written as,

\begin{widetext}
\begin{eqnarray}
{\cal F}({\it q},{\it E}_{c})={\it i}\frac{{\it M}^{2}}{2{\it E}_{c}k_{c}}
[{\it F}^{L}_{D}({\it q}) + {\it F}^{L}_{X}({\it Q})],
\end{eqnarray}
\begin{eqnarray}
{\it F}^{L}_{D}({\it q}) = \sum_{i}\delta_{L, L({\it i})}\lbrace\tau_{0}. 
\tau_{1}\rbrace^{\it I_{i}}{\it f}^{i}(q),
\end{eqnarray}
\begin{eqnarray}
{\it F}^{L}_{X}({\it Q}) =(-1)^{\it T} \sum_{i}B_{L({\it i}),L}\lbrace\tau_{0}. 
\tau_{1}\rbrace^{\it I_{i}}{\it f}^{i}(Q),
\end{eqnarray}

\begin{eqnarray}
{\it f}^{i}(q)& \equiv &\frac{{\it g}^{2}_{i}}{q^{2}+m^{2}_{i}}
{\left( \frac{\Lambda^{2}_{i}}{{q}^{2}+{\Lambda^{2}_{i}}}\right)}^{2} 
-{\it i}\frac{\overline{\it g}^{2}_{i}}{Q^{2}+\overline{m^{2}_{i}}}
{\left(\frac{\overline\Lambda^{2}_{i}}{{q}^{2}+{\overline\Lambda^{2}_{i}}}\right)}^{2}.
\end{eqnarray}
\end{widetext}
Here ${\it E}_{c}$ is the total energy in the NN center of mass system. 
Note that ${\it f}^{i}$ depends only on the magnitude of the three momentum 
transfer and the expressions are used to fit NN-scattering amplitude at 
laboratory energy. The full parametrizations are frame out 
in Refs. \cite{witz85,mur87}.

\section{Nucleon-Nucleus Optical Potential}

The Dirac optical potential ${\it U}_{opt}(q, E)$ can be written as,
\begin{eqnarray}
{\it U}_{opt}(q, E) = \frac{-4\pi ip}{M}\langle\psi\vert\sum_{n=1}^{A} 
\exp{(iq.x(n))}{\cal F}(q, E; n)\vert\psi\rangle
\end{eqnarray}
where ${\cal F}$ is the scattering operator, {\it p} is the momentum of the 
projectiles in the nucleon-nucleus center of mass frame, $\vert\psi\rangle$ 
is the nuclear ground state wave function for A-particle, {\it q} is the 
momentum transfer and {\it E} is the collision energy for a stationary 
target (nucleus) and incident projectile (proton). In the present 
calculation the nuclear recoil energy is neglected because of elastic 
scattering. The operator ${\cal F}(q, E; n)$ describe the scattering of the 
projectile from target nucleon 'n' without separation into direct and 
exchange terms. Let us define the nuclear ground state by a Dirac-Hartree 
wave function \cite{dock87} and the incident projectile wave function 
as ${\cal U}(x)$, then the optical potential on incident wave projected 
to the co-ordinate space can be written as,

\begin{widetext}
\begin{eqnarray}
\langle x \vert{\it U}_{opt} \vert{\cal U}_{0}\rangle&=& 
\frac{-4\pi ip}{M}\langle\psi\vert\sum_{\alpha}^{occ}
\int d^{3}y^{'} d^{3}y d^{3}x^{'}{\cal U}_{\alpha}(y^{'})
\times\left\lbrace\langle xy^{'}\vert{\it t}(E)\vert x^{'}y\rangle
+(-1)^{T}\langle y^{'}x\vert{\it t}(E)\vert x^{'}y 
\rangle\right\rbrace {\cal U}_{0} (x^{'}){\cal U}_{\alpha} (y).
\end{eqnarray}

The antisymmetrised matrix element of ${\it t}(E)$ in coordinate space is the 
Fourier transforms \cite{dock87} of the matrix element in the momentum space 
co-ordinate and is written as,

\begin{eqnarray}
\langle x \vert{\it U}_{opt} \vert{\cal U}_{0}\rangle&=& 
\frac{-4\pi ip}{M}\sum_{L}\int d^{3}x^{'} 
\left[ \rho^{L}(x^{'})t^{L}_{D}(\vert x-x^{'} \vert ; E)\right] 
\lambda^{L}{\cal U}_{0}(x)\nonumber\\
&& - \frac{-4\pi ip}{M}\sum_{L}\int d^{3}x^{'} 
\left[\rho^{L}(x^{'},x)t^{L}_{X}(\vert x-x^{'} \vert ; E)\right] 
\lambda^{L}{\cal U}_{0}(x^{'}),
\end{eqnarray}
\end{widetext}
where
\begin{eqnarray}
t^{L}_{D}(\vert x \vert; E)\equiv \int \frac{d^3q}{(2\pi)^3} 
t^{L}_{D}(q,E) e^{-iqx},\nonumber\\
\end{eqnarray}
with
\begin{eqnarray}
t^{L}_{D}(\vert q \vert; E)\equiv (\frac{iM^{2}}{2E_{c}k_{c}})F^{L}_{D}(q),\nonumber\\
\end{eqnarray}
and similarly for the exchange part $t^{L}_{D}(\vert Q \vert; E)$. The nuclear density 
is defined by a simple expression similar to the equation of RMF and E-RMF density, 

\begin{eqnarray}
\rho^{L}(x, x^{'})\equiv\sum_{\alpha}^{occ^{'}}\overline{{\cal U}_{\alpha}}\lambda^{L}{\cal U}_{\alpha},
\rho^{L}(x)\equiv\rho^{L}(x^{'},x).\nonumber\\
\end{eqnarray}
The prime stands for occupied states, i.e., sum over target protons 
(pp-amplitude) and target neutrons (pn-amplitude)  used. The first term 
in the  Eqn. (22) defines the direct optical potential,

\begin{eqnarray}
U^{L}_{D}(r, E) = \frac{-4\pi ip}{M}\int d^{3}x^{'} \rho^{L}(x^{'})t^{L}_{D}
(\vert x-x^{'} \vert ; E).
\end{eqnarray}
The nonlocal second term is treated in nonlocal density approximation 
\cite{bri77}, which contains plane wave status for incident and bound 
nucleons. We replaced the exchange integral with local potential by,

\begin{widetext}
\begin{eqnarray}
U^{L}_{X}(r, E) = \frac{-4\pi ip}{M}\int d^{3}x^{'} \rho^{L}(x^{'}, x)
t^{L}_{D}(\vert x-x^{'} \vert ; E)j_{0}(p(\vert x-x^{'}\vert ),
\end{eqnarray}
where $j_{0}$ is the spherical Bessel-function. The off diagonal one body 
density is approximated by the local density which result as,

\begin{eqnarray}
\rho_{L}(x^{'},x)\approx \rho^{L}(1/2(x+{x^{'}))(\frac{3}{sk_{f}})j_{1}(sk_{f})},
\end{eqnarray}
\end{widetext}
with $s\equiv\vert x-x^{'}\vert$ and $k_{f}$ is related to the nuclear baryon 
density by $\rho_{B}(1/2(x+x^{'}))=2k^3_{f}/3\pi ^{2}$. Now the optical 
potential have the form,

\begin{eqnarray}
U_{opt}=U_{S}+\gamma^{0}U^{V}-2i\alpha.\widehat{r}U^{T},
\end{eqnarray}
where
\begin{eqnarray}
U^{L}\equiv U^{L}(r, E)=U^{L}_{D}(r, E)+U^{L}_{X}(r, E).
\end{eqnarray}
As the tensor contributions are small, by neglecting these, the Dirac 
equation for projectile has precisely the similar form as in RMF and 
E-RMF equation. By taking the Fourier Transform of this equation, we get the 
optical potential as, 



\begin{widetext}
\begin{eqnarray}
\int\frac{d^{3}q}{(\pi)^{3}}\exp{(iq.x)}f(q) = 
\frac{g^{2}}{4 \pi}\frac{\Lambda ^{2}}{\Lambda ^{2}-m^{2}}
\left\lbrace\frac{\Lambda ^{2}}{\Lambda ^{2}-m^{2}}
\frac {e^{-mr}-e^{-\Lambda r}}{r}-\frac{\Lambda}{2}e^{-\Lambda r}\right\rbrace.
\end{eqnarray}
\noindent
This equation includes all meson exchanges (except the pseudoscalar meson) with 
derivative coupling, which is written in the form,

\begin{eqnarray}
\int \frac{d^{3}q}{(2\pi)^{3}}\exp{(iq.x)}f(q) \frac{q^{2}}{4M^{2}}
= \frac{\Lambda^{2}}{4M^{2}} \frac{g^{2}}{4 \pi} \frac{\Lambda ^{2}}{\Lambda ^{2}-m^{2}}
\left\lbrace\frac{m^{2}}{\Lambda^{2}-m^{2}}
\frac{e^{-\Lambda r}-e^{-mr}}{r}+\frac{\Lambda}{2} e^{-\Lambda r}\right\rbrace.
\end{eqnarray}
\end{widetext}
\noindent
The optical potential is modified by Pauli blocking factor 
\cite{ser84,mann84,mac85,har86,ser87} ${\it a(E)}$ with local density 
approximation as follows,

\begin{eqnarray}
{\it U}^{L} (r, E)\longrightarrow\left[ 1-a(E)(\frac{\rho_{B} (r)}{\rho_{0}})^{2/3}\right] 
{\it U}^{L}(r,E).
\end{eqnarray}

\noindent
Here $\rho_{B}$ is the local baryon density of the target and $\rho_{0}$ 
is the nuclear matter density at saturation. The approximation depends on 
$\rho_{B}^{2/3}$, which agree with phase-space arguments based on isotropic 
scattering. The detail about the Pauli blocking factor is given in  Ref. 
\cite{mur87}. To solve the scattering state Dirac equation, the wave function 
is separated into two components (upper and lower) and this equation is 
expressed as two coupled first order differential equations. Elimination of 
the lower component leads to a single second order differential equation with
spin-orbit as well as both local and nonlocal potential. The 
nonlocal Darwin potential can be separated by rewriting the upper component of the wave 
function,
${\cal A}^{1/2}(r, E){\cal U}(x)$  and

\begin{eqnarray}
{\cal A}(r, E) \equiv 1+\frac{U^{S}(r, E)-U^{V}(r, E)}{E+M}.
\end{eqnarray}
After some algebra, the equation can be written as,
\begin{widetext}
\begin{eqnarray}
(-\nabla^{2} + V_{cent}+V_{so}\sigma.L +V_{Darwin}) u(x)=(E^{2}-M^{2})u(x),
\end{eqnarray}
\end{widetext}
where the energy-dependent optical potentials are

\begin{eqnarray}
V_{cent} (r, E)= 2MU^{S}+2EU^{V}+(U^{S})^{2}-(U^{V})^{2},\\
V_{so}(r, E)=-\frac{1}{r}\frac{B^{'}}{B},\\
and\nonumber\\
V_{Darwin}=\frac{3}{4}(\frac{B^{'}}{B})^{2}-\frac{1}{r}\frac{B^{'}}{B}-\frac{1}{2}
\frac{B^{''}}{B}.
\end{eqnarray}
Since the two component Dirac wave functions are eigenstate of ${\sigma .L}$, so 
by taking the second derivative of the function we can solve easily using 
Numerov algorithm \cite{koon86,koon95}. Note that ${{\cal U}(x)}$ is not 
equal to the upper component wave function in the region of the potential 
but when ${A}(r, E)\longrightarrow$1, as 
$r\longrightarrow \infty$ and ${\cal{U}}$ has the same asymptotic behavior the 
wave function at large ${r}$. Thus the correct boundary condition is imposed 
by matching   ${\cal{U}}$ to the form of Coulomb scattering solution incident 
in the z-direction \cite{thy68}.

\begin{widetext}
\begin{eqnarray}
\psi(r)&\propto_{r \longrightarrow \infty}&
\left\lbrace \exp{i[pz-\eta ln 2pr sin^{2}\theta /2]}
\left[ 1- \frac {\eta^{2}}{2 i p r sin^{2}\theta /2}\right] 
\right\rbrace \chi_{inc}+\left\lbrace\frac {\exp{i[pr-\eta ln 2pr]}}{r}
\left[ A(\theta)+ B(\theta) \sigma . \widehat{n}\right] \right\rbrace 
\chi_{inc}, 
\end{eqnarray}
\end{widetext}
\noindent
with $E=\sqrt{p^{2}+M^{2}}$, $\chi_{inc}$ is a two-component Pauli spinor, 
$\theta$ is the scattering angle, $n$ is the normal to the scattering plane 
and $\eta \equiv Ze^{2}/p^{2}$ with Z is the nuclear charge. The scattering 
observables like differential scattering cross-section 
($\frac{d\sigma}{d\Omega}$) and other quantities, like optical potential 
($U_{opt}$), analyzing power ($A_y$) and spin observables ($Q-$value) are 
easily determined from the scattering amplitude, which are written as,

\begin{eqnarray}
\frac{d\sigma}{d\Omega}\equiv\vert A(\theta)\vert ^{2}+\vert B(\theta)\vert ^{2},\\
A_{y}\equiv\frac{2Re[A^{*}(\theta)B(\theta)]}{d\sigma /d\Omega},\\
Q\equiv\frac{2Im[A(\theta)B^{*}(\theta)]}{d\sigma /d\Omega}. 
\end{eqnarray}
\begin{figure}[ht]
\vspace{1.0cm}
\begin{center}
\includegraphics[width=1.0\columnwidth]{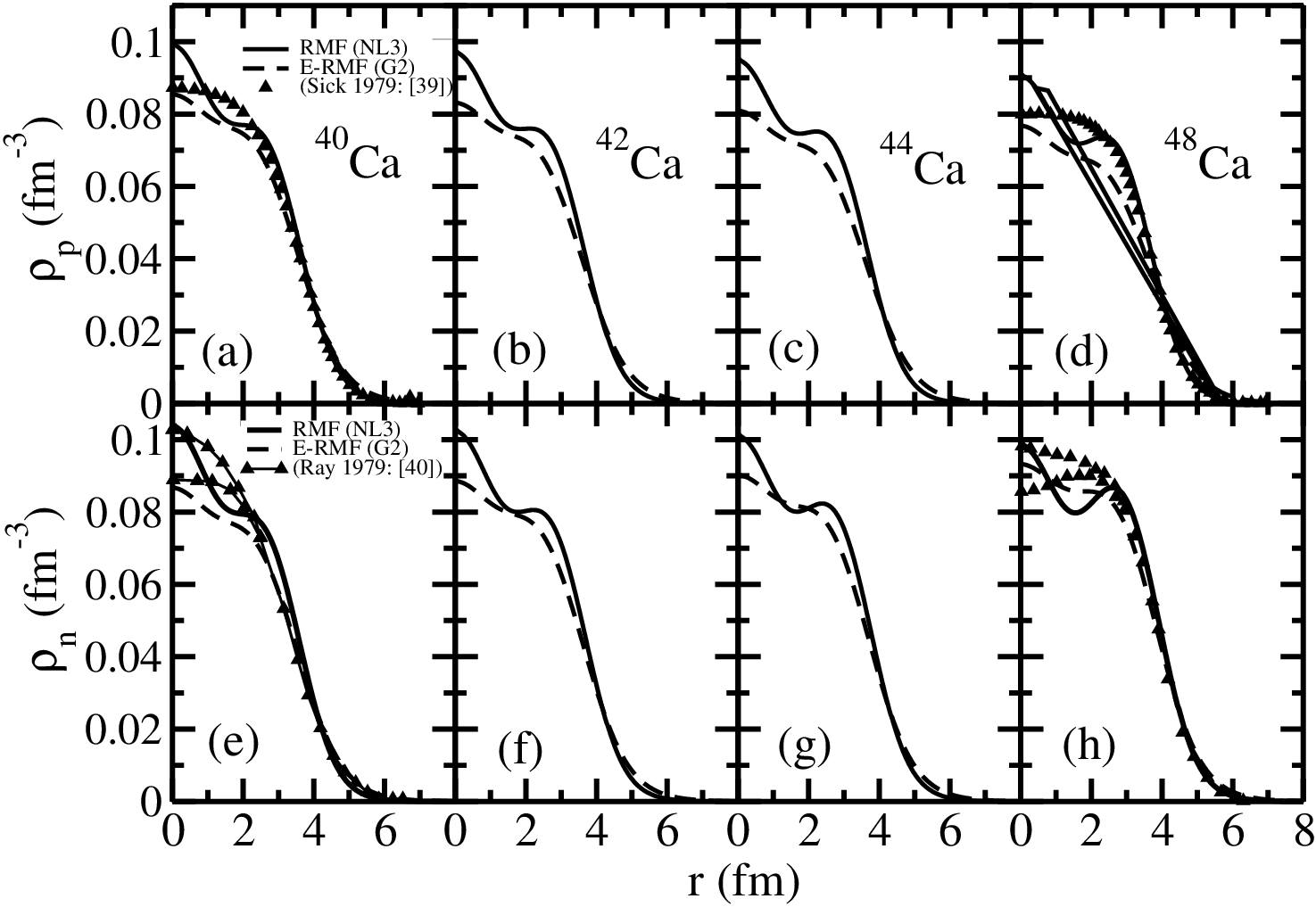}
\caption{The proton $\rho_p$ (upper panel) and neutron $\rho_n$ (lower panel) density distribution for
$^{40,42,44,48}Ca$ obtained from RMF (NL3) and E-RMF (G2) parameter sets. The experimental \cite{sick79}
$\rho_p$ and deduced \cite{ray79a}  $\rho_n$ for $^{40,48}Ca$ are also compared. 
}
\end{center}
\label{Fig. 1}
\end{figure}

\begin{figure}[ht]
\vspace{1.0cm}
\begin{center}
\includegraphics[width=0.70\columnwidth]{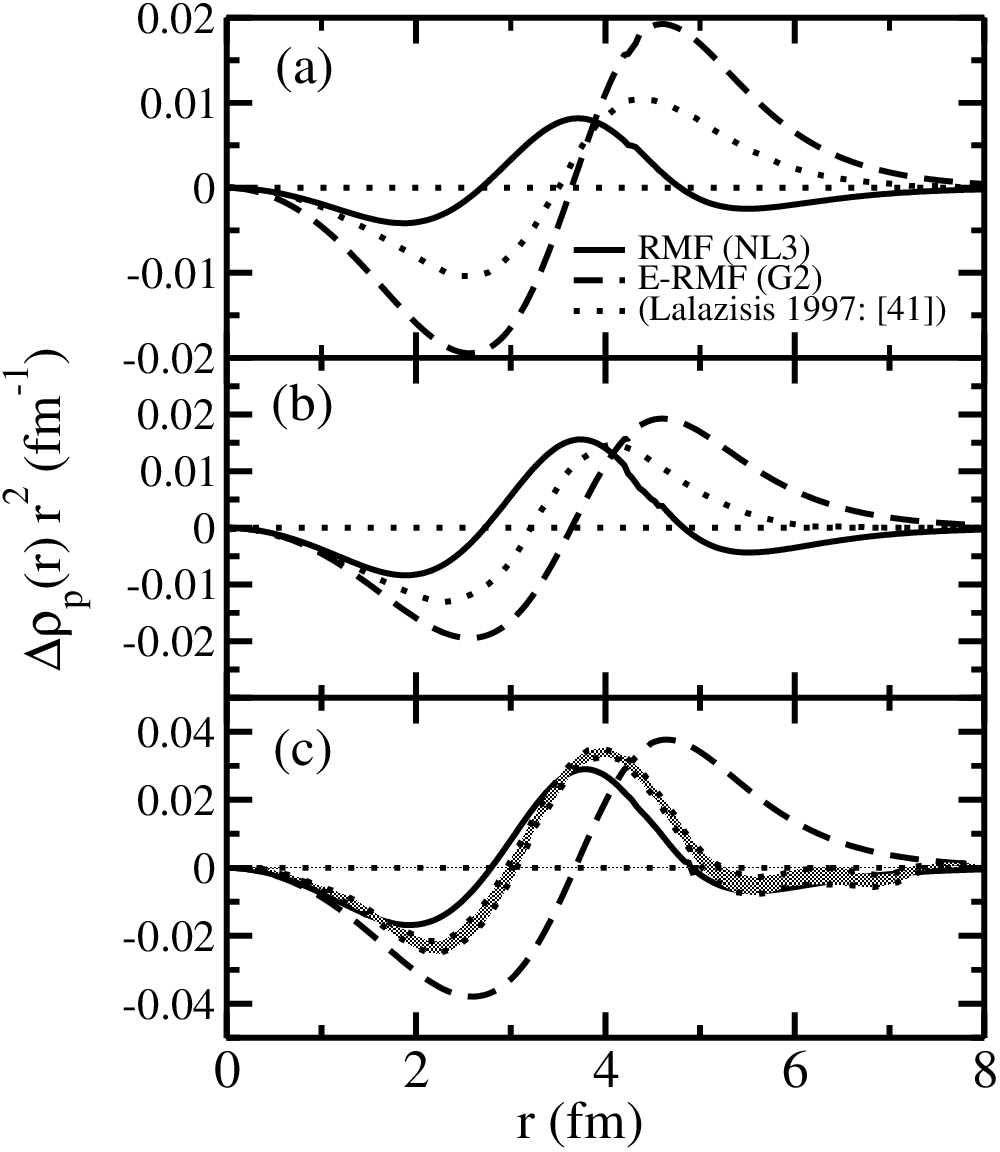}
\caption{The relative isotopic proton density differences $\Delta\rho_p(r)$ for 
$\rho_p(^{42}Ca)-\rho_p(^{40}Ca)$,
$\rho_p(^{44}Ca)-\rho_p(^{40}Ca)$,
and $\rho_p(^{48}Ca)-\rho_p(^{40}Ca)$ obtained from RMF (NL3) and E-RMF (G2)  are compared with the
data  \cite{lala27a} in (a), (b) and (c), respectively.
}
\end{center}
\label{Fig. 2a}
\end{figure}

\begin{figure}[ht]
\vspace{1.0cm}
\begin{center}
\includegraphics[width=1.0\columnwidth]{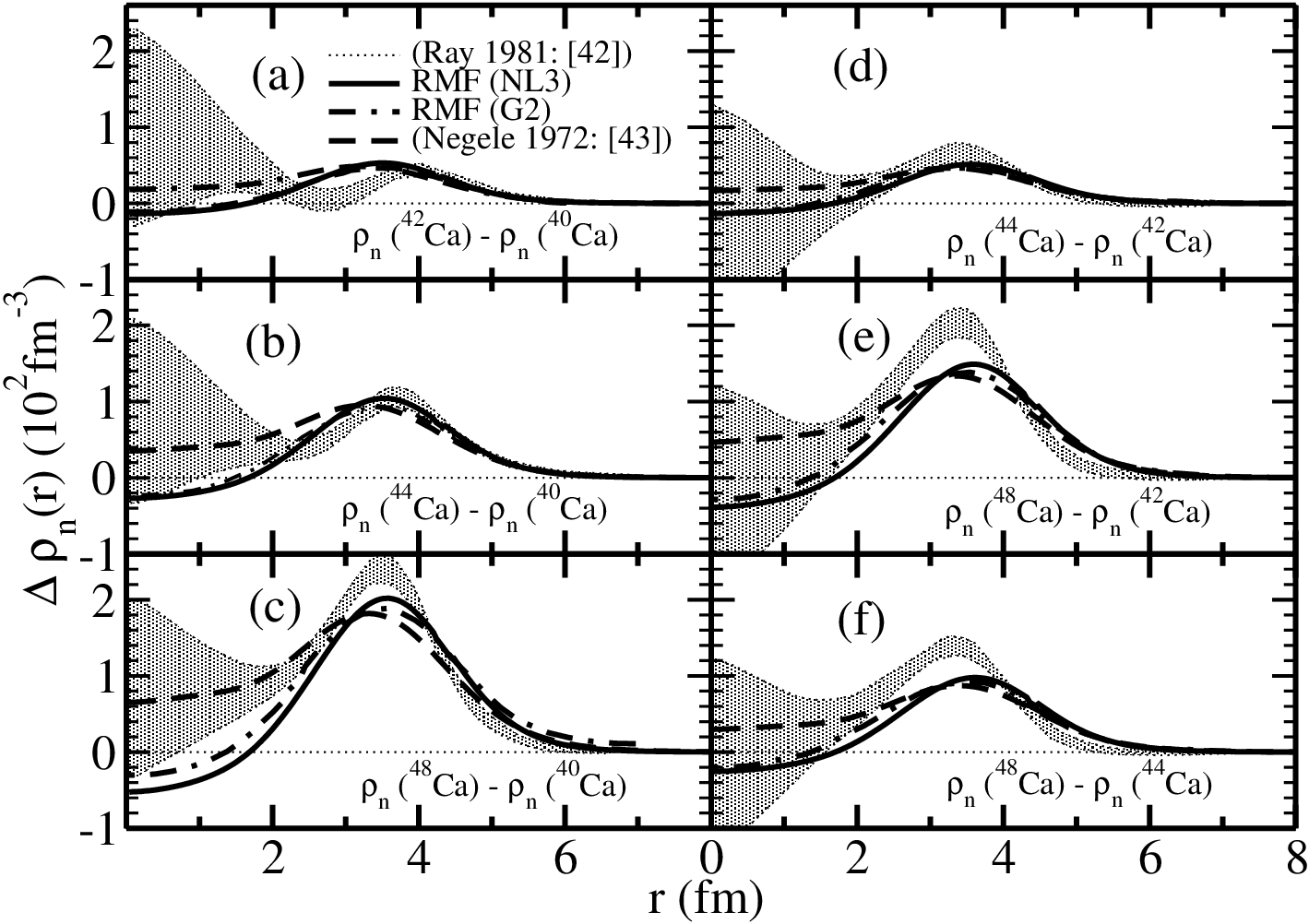}
\caption{The relative isotopic neutron density differences $\Delta\rho_n(r)$ 
for $\rho_n(^{42}Ca)-\rho_n(^{40}Ca)$,
$\rho_n(^{44}Ca)-\rho_n(^{40}Ca)$,
$\rho_n(^{48}Ca)-\rho_n(^{40}Ca)$,
$\rho_n(^{44}Ca)-\rho_n(^{42}Ca)$,
$\rho_n(^{48}Ca)-\rho_n(^{42}Ca)$,
and $\rho_n(^{48}Ca)-\rho_n(^{44}Ca)$.
The RMF (NL3) and E-RMF (G2) $\Delta\rho_n(r)$ are compared with the 
density-matrix-expansion (DME) data \cite{dme81} and the uncertainty deduced neutron 
difference \cite{ray81}.
}
\end{center}
\label{Fig. 2b}
\end{figure}

\section{Details of the Calculations and Results}

First we calculate both the scalar and vector parts of the neutrons and 
protons density  distribution for $^{40,42,44,48}$Ca from the RMF (NL3) 
and E-RMF (G2) formalisms \cite{patra01a}. Then evaluate the scattering 
observables using these densities in the RIA frame-work\cite{horo90}, 
which involves the following two steps: 
(i) we generate the complex NN-interaction from the 
Lorentz invariant matrix ${\cal F}^L(q,E)$ as defined in Eq. (2). Then 
the interaction is folded with the ground state target nuclear density 
for both the RLF \cite{witz85,mur87} and MRW parameters \cite{neil83} separately and obtained 
the nucleon-nucleus complex optical potential $U_{opt}(q,E)$. It is to be 
noted that the pairing interaction has been taken into account using the Pauli blocking 
approximation. Here, the Pauli blocking enters through the intermediate 
states of the {\it t}-matrix formalism, which has geometrical effects 
on the optical potential, (ii) we solve the wave 
function of the scattering state utilising the optical potential prepared 
in the first step by the well known Numerov algorithm \cite{koon86}. The 
result is approximated with the non-relativistic Coulomb scattering for 
a wide range of radial component which yields the scattering amplitude 
and other observables \cite{thy68}. By comparing our calculations with the
available experimental data, we examine the validity of our RIA predictions for 
describing  $\frac{d\sigma}{d\Omega}$, $A_y$ and Q-values which are 
presented in Figures $1-11$.

\begin{figure}[ht]
\vspace{1.0cm}
\begin{center}
\includegraphics[width=1.0\columnwidth]{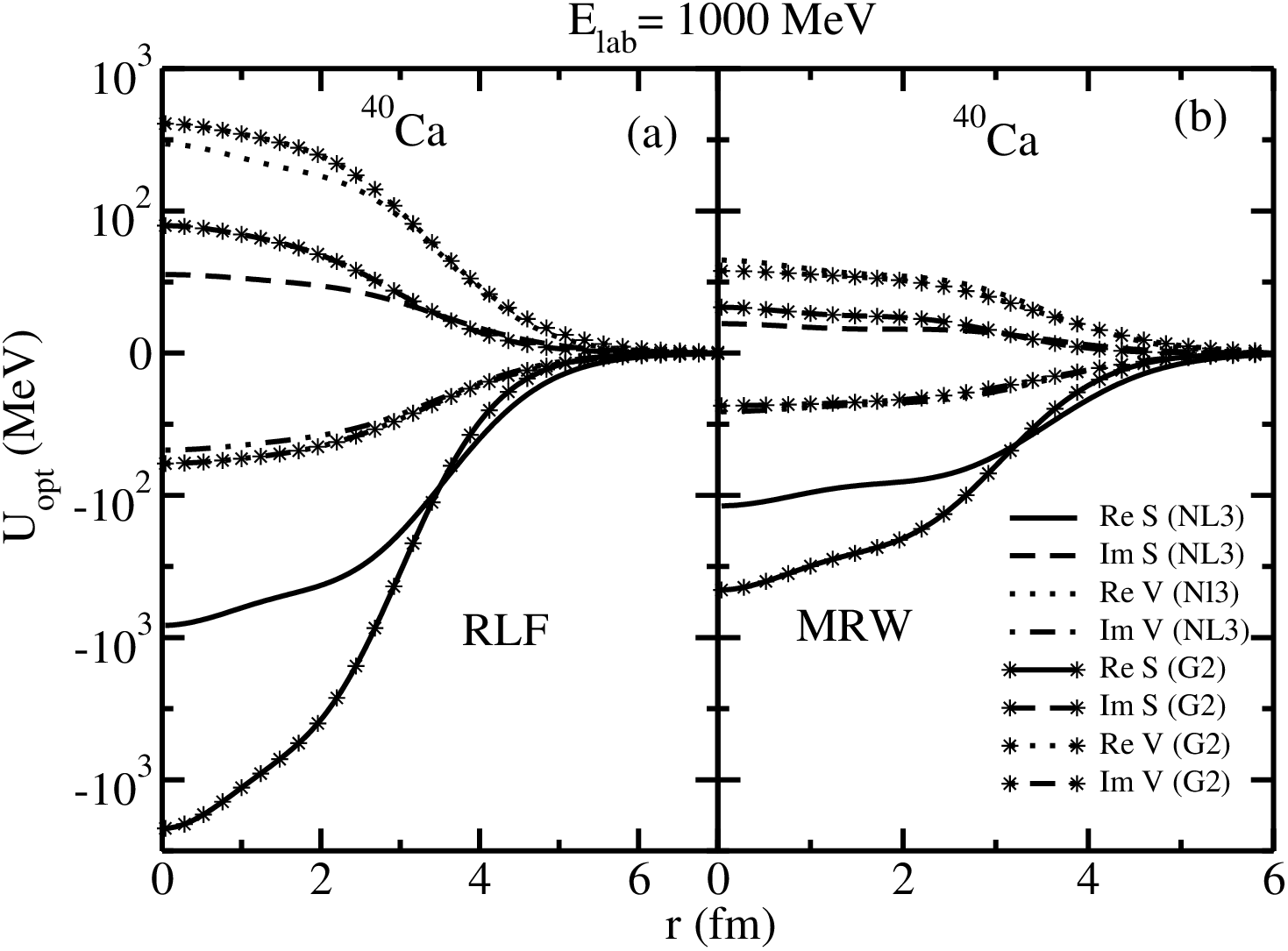}
\caption{The real (Re) and imaginary (Im) part of the scalar S and vector 
V  Dirac optical potential is plotted as a function of the nuclear radius 
for $p+^{40}Ca$ system using RMF (NL3) and E-RMF (G2) densities. (a) is for 
RLF and  (b) is for MRW parametrization. The energy of the projectile proton 
is $E_{lab}=1000$ MeV.
}
\end{center}
\label{Fig. 3}
\end{figure}

\begin{figure}[ht]
\vspace{1.0cm}
\begin{center}
\includegraphics[width=1.0\columnwidth]{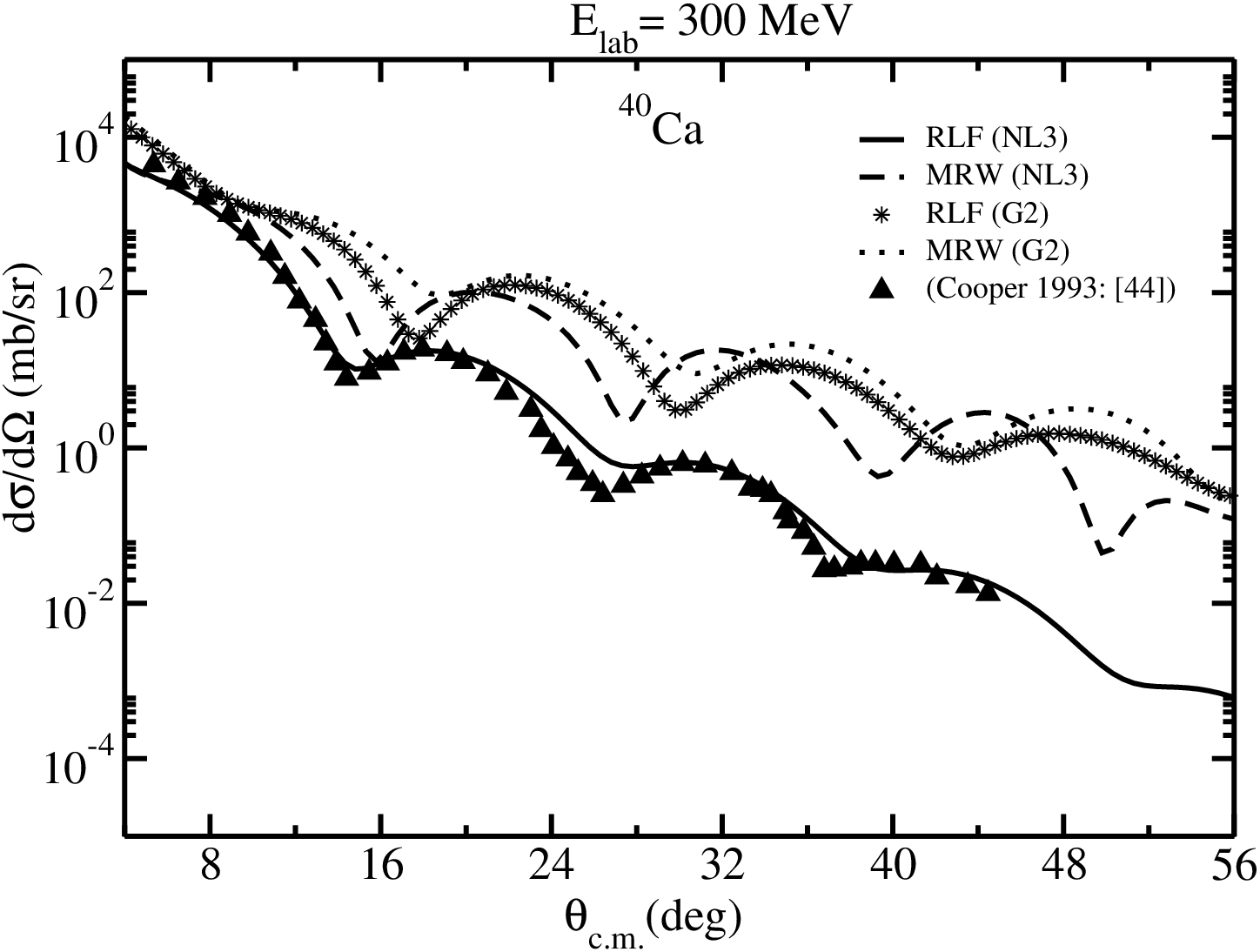}
\caption{The elastic differential scattering cross-section
($\frac{d\sigma}{d\Omega}$) as a function of scattering angle
$\theta_{c.m.}$(deg) for $^{40}Ca$ using both RLF and MRW
parametrizations at $E_{lab}=300$ MeV. Triangles are the experimental
data  \cite{coop93}.
}
\end{center}
\label{Fig. 4}
\end{figure}

\begin{figure}[ht]
\vspace{1.0cm}
\begin{center}
\includegraphics[width=1.0\columnwidth]{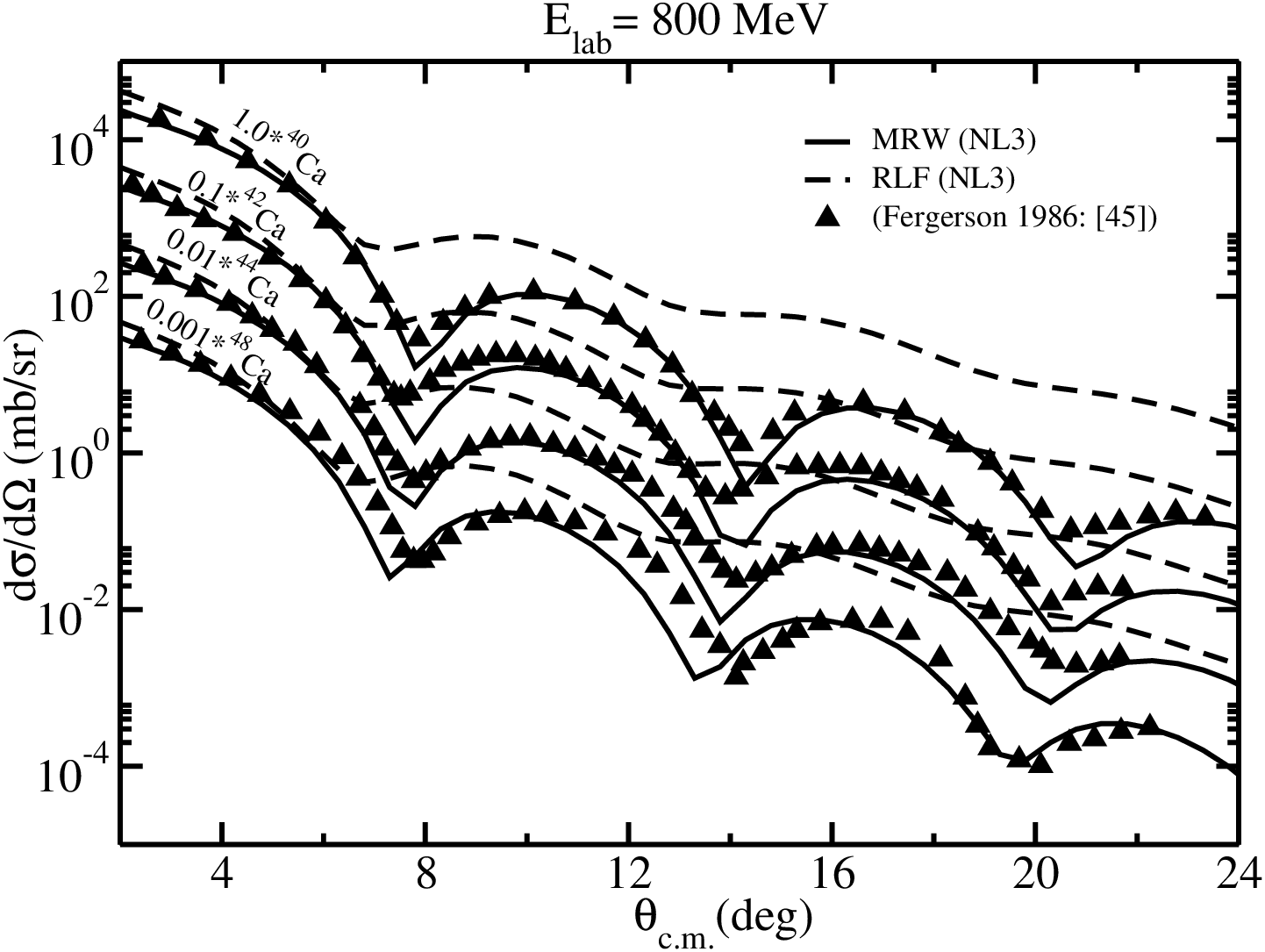}
\caption{Same as Fig. 5 for  $^{40,42,44,48}Ca$  at $E_{lab}=800$ MeV.
The experimental data are taken from  \cite{ferg86}.
}
\end{center}
\label{Fig. 5}
\end{figure}

\begin{figure}[ht]
\vspace{1.0cm}
\begin{center}
\includegraphics[width=1.0\columnwidth]{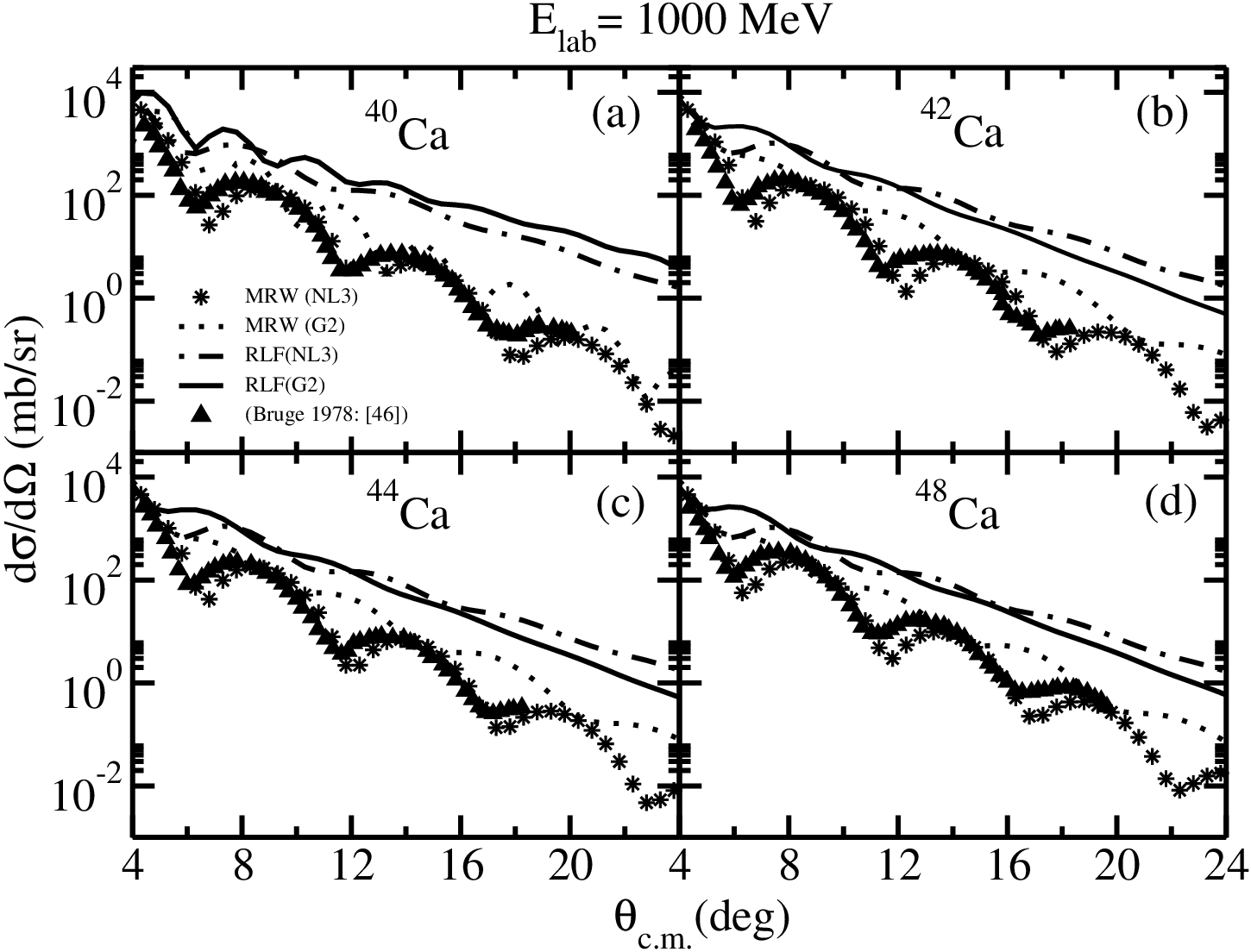}
\caption{
Same as Fig. 6  at $E_{lab}=1000$ MeV.
The experimental data are taken from  \cite{expt78}.
}
\end{center}
\label{Fig. 6}
\end{figure}

\subsection{The neutron and proton densities}

In Fig. 1, we have plotted the proton $\rho_p$ and neutron $\rho_n$ density 
distribution for $^{40,42,44,48}Ca$ using NL3 and G2 parameter sets within RMF and E-RMF 
formalisms. From the figure, we note that, there is a very small 
difference in the densities for NL3 and G2 parameter sets. However, a careful 
inspection shows a small enhancement in central density (0-1.6 fm) for NL3 
set. On the other hand the densities obtained from G2 is elongated to a larger 
distance towards the tail region and this nominal 
difference has significant role to play in the scattering phenomena, 
which is explained later on. Further, the agreement of $\rho_p$ with
the experiment \cite{sick79} and $\rho_n$ with the deduced data \cite{ray79a} for NL3 set is 
slightly better than that of G2. Explicitly, it is worth
mentioning that the $\rho_p$ (NL3) matches with the data even at the central region,
whereas the $\rho_p$ of G2 under-estimates through out the density plot.  

A microscopic investigation of Fig. 1 shows a change in $\rho_p(r)$, $\rho_n(r)$, 
i.e. the area covered by the proton and neutron densities gradually increases with the mass number
in an isotopic chain. From the $\rho_p(r)$ and $\rho_n(r)$, we estimate the possible 
relative isotopic density difference $\Delta\rho(r)$ for RMF 
(NL3) and E-RMF (G2) parameter sets (see Figs. 2 and 3). The calculated
$\Delta\rho_p(r)$ are compared with the experimental data \cite{lala27a} in Fig. 2. The
measured data of $\Delta\rho_p(r)$ lies in between the prediction of NL3 and G2 values
as shown in Fig. 2. Comparing   $\rho_p(^{42}Ca)-\rho_p(^{40}Ca)$, $\rho_p(^{44}Ca)-\rho_p(^{40}Ca)$,
and $\rho_p(^{48}Ca)-\rho_p(^{40}Ca)$ of Fig.2 [(a), (b) and (c)], we notice a better agreement
of NL3 values over G2 with respect to experimental measurement in the 
isotopic chain which subsequently reflects in the results of scattering observables.
 
The relative isotopic density difference for neutron $\Delta\rho_n(r)$ is 
compared in Fig. 3 with the deduced neutron density difference data \cite{ray81} 
and the density-matrix-expansion prediction \cite{dme81}. 
The predicted results with RMF (NL3) agree well only for the double closed shell nuclei 
$^{40}Ca$ and $^{48}Ca$. But in case of E-RMF (G2) we get excellent
match with the deduced $\Delta\rho_n(r)$ for the considered isotopic 
chain. There is a peak appears in $\Delta \rho_n(r)$ at radial range 
$r\sim 3.4-3.8 fm$ and this peak slightly shifted towards the center 
with the increase of neutron number. Although $\Delta\rho_n(r)$ for G2 set gives better agreement
with  the deduced values, the use of NL3 set in the RIA formalism works well for the scattering 
observables (shown later).

\begin{figure}[ht]
\vspace{1.0cm}
\begin{center}
\includegraphics[width=1.0\columnwidth]{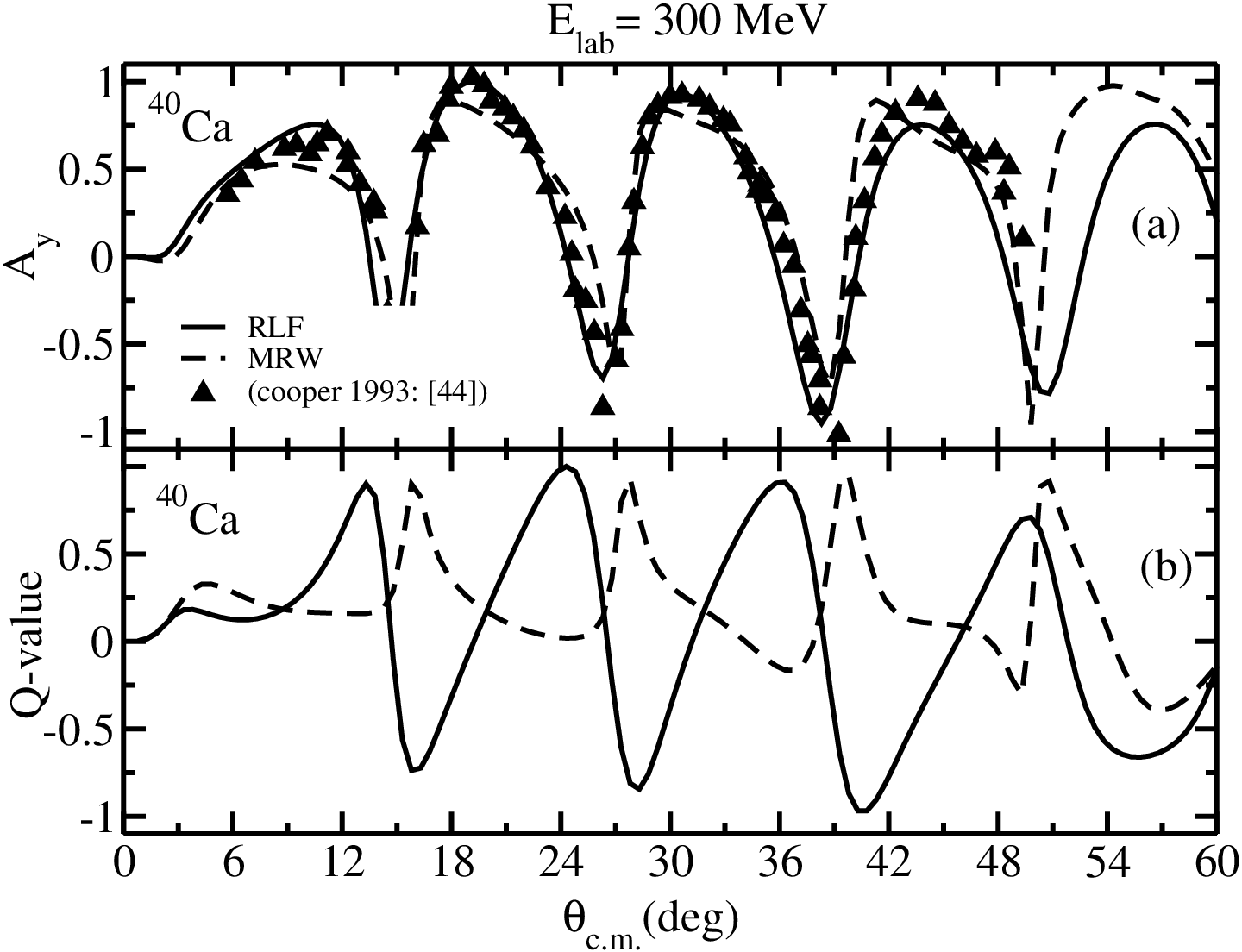}
\caption{Analyzing power $A_y$ and Q-Values
(spin observables) as a function of scattering angle $\theta_{c.m.}$(deg)
for $^{40}Ca$ at $E_{lab}=300$ MeV. The experimental data are taken from  
\cite{coop93}.
}
\end{center}
\label{Fig. 7}
\end{figure}

\begin{figure}[ht]
\vspace{1.0cm}
\begin{center}
\includegraphics[width=1.0\columnwidth]{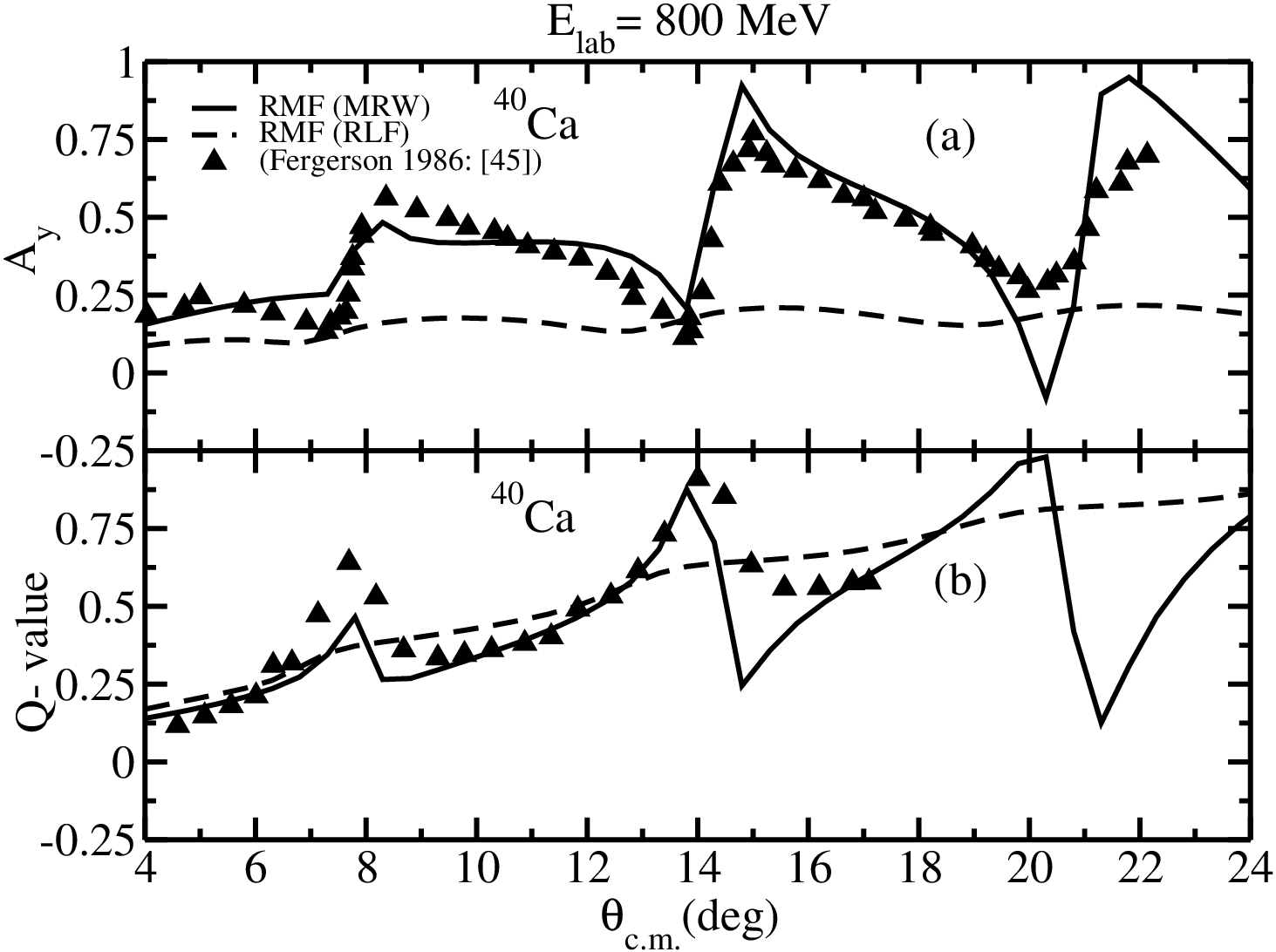}
\caption{Same as Fig. 8, but at $E_{lab}=800$ MeV. 
The experimental data are taken from  \cite{ferg86}. 
}
\end{center}
\label{Fig. 8}
\end{figure}

\begin{figure}[ht]
\vspace{1.0cm}
\begin{center}
\includegraphics[width=1.0\columnwidth]{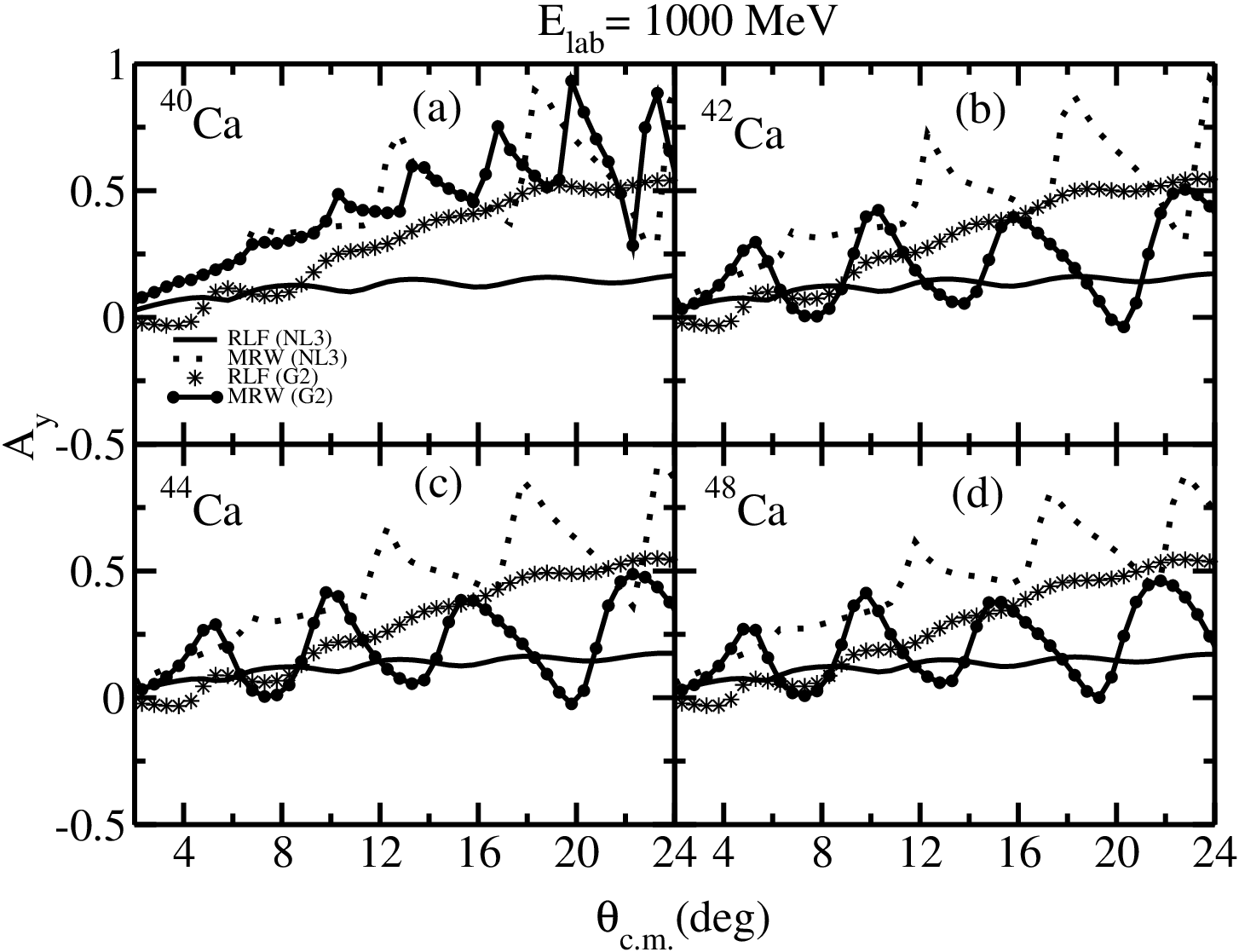}
\caption{Analyzing power $A_y$ as a function
of scattering angle $\theta_{c.m.}$(deg) for $^{40,42,44,48}Ca$
at $E_{lab}=1000$ MeV.
}
\end{center}
\label{Fig. 9}
\end{figure}

\begin{figure}[ht]
\vspace{1.0cm}
\begin{center}
\includegraphics[width=1.0\columnwidth]{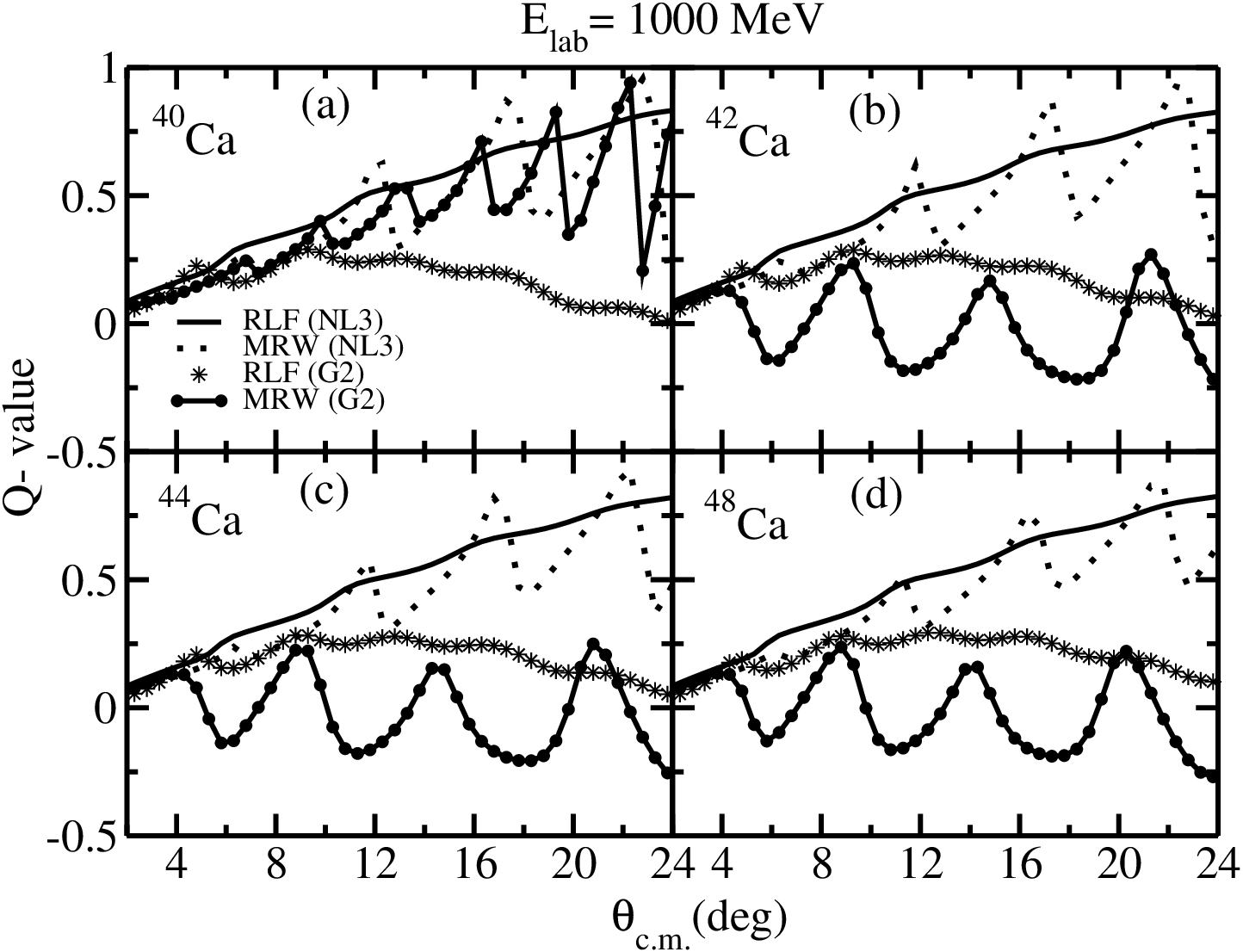}
\caption{The spin observable $Q-$value as a function of scattering
angle $\theta_{c.m.}$(deg) for $^{40,42,44,48}Ca$ at $E_{lab}= 1000$ MeV.
}
\end{center}
\label{Fig. 10}
\end{figure}

\subsection{Optical potential}
  
With the densities in hand, we calculate the optical potential $U_{pot}$ for 
$^{40,42,44,48}Ca$ by folding the density matrix with the NN scattering 
amplitude of the proton projectile for 300, 800 and 1000 MeV. The $U_{pot}$ 
is a complex function which constitute both real and imaginary 
part for both the scalar and the vector potentials. In Fig. 4, we present the $U_{pot}$ 
for $\overrightarrow{p}+^{40}Ca$ at laboratory energy $E_{lab}=1000$ MeV 
as a representative case. We also examine the $U_{pot}$ for other Ca isotopes 
and find similar trends with $\overrightarrow{p}+^{40}Ca$. In other words, 
we do not get any significant difference in the optical potential with 
the increase of neutron number. Similar to the density distribution in NL3 
and G2 (Fig. 1), here we find a difference in $U_{opt}(q,E)$ between the 
RLF and  MRW parametrizations. The evaluation methods of the optical 
potentials using RLF or MRW (see Fig. 4) are somewhat different from each other, which 
is given in {\bf Appendix A} \cite{horo90}, which is responsible 
for the use of the different parametrizations at various ranges of incident energies. 
For example, the RLF parameters used here are from Refs. \cite{witz85,mur87} 
which are computed for energies up to 400 MeV and are therefore suitable for 
lower $E_{lab}$ whereas the MRW is better for the  
higher values which will be discussed in the coming sections. Further, the 
$U_{opt}(q,E)$ values from either RLF or MRW, differs significantly depending on 
the NL3 or G2 force parameters. That means, the optical potential is not only sensitive to 
RLF or  MRW but also to the use of NL3 or G2 densities. 
Investigating the figure, it is clear that 
the extreme values of the magnitude of real and imaginary part of the scalar 
potential are -382.9 and 110.6 MeV for RLF (NL3) and -372.4 and 177.8 MeV 
for RLF (G2) respectively. The same values for the MRW parametrization are -217.7 and 
40.2 MeV with the NL3 and -333.8 and 61.7 MeV with the G2 sets. In case of 
the vector potential, the extreme values for the real and imaginary parts are 
293.0 and -136.0 MeV for RLF (NL3) and 319.7 and -157.5 MeV for RLF (G2) 
but with MRW parametrization these appear at 124.1 and -82.3 MeV 
with the NL3 and 115.5 and -77.1 MeV with the G2. From these variations in the magnitude 
of scalar and vector potentials, it is clear that the predicted results 
not only depend on the input target density, but also sensitive 
to the kinematics of the reaction dynamics. A further analysis of the 
results for the optical potential with RLF, it is noticed that the 
$U_{opt}$ value extends for a larger distance than MRW. For example, 
with RLF the central part of $U_{opt}$ is more expanded than MRW and 
ended at $r\sim 5 fm$, whereas the $U_{opt}$ persists till 
$r\sim 6 fm$. It is important to point out that the lack of the availability 
of experimental data for optical potential, we are unable to 
justify the capability of parametrizations at different energies. 
We also repeat the calculations without Pauli blocking and found almost 
identical results for optical potential at $E_{lab}\sim$ 300, 800 and 1000 MeV.
The effects of RLF and MRW parametrizations are presented in the next
subsections during the discussion of scattering observables.

\subsection{Differential scattering cross-section} 

Evaluation of the differential elastic scattering cross-section 
$\frac{d\sigma}{d\Omega}$, defined in Eqn. (40) is crucial to study 
the scattering phenomena. The results of our calculation for 
$\overrightarrow{p}+^{40}Ca$ and $\overrightarrow{p}+ ^{40,42,44,48}Ca$ 
systems at incident energies 300, 800 and 1000 MeV, respectively are displayed 
in Figs. 5, 6 and 7 along with the available experimental data \cite{coop93, ferg86, expt78}.
As it is stated earlier that the RIA prediction with the NL3 density 
is better to the choice of G2 for all the angular distributions, irrespective of
the use of RLF or MRW parametrizations. Again considering the energy of the projectile, 
the RLF predictions best fit to the data for $E_{lab}\leq 400$ MeV
(see Fig. 5). However, results obtained from the MRW parametrization is better for 
higher incident energies (Figs. 6 and 7) ($E_{lab} > 400$ MeV) \cite{neil83,horo90}.
This result shows a fundamental difference between the RLF and MRW parametrization 
depending upon the incident energy ranges. Perhaps due to this reason, the explicit off 
shell behavior of RLF and MRW is drastically affecting the scattering predictions. 
Similarly for the optical potential the results are insensitive to the Pauli blocking.

\subsection{Analyzing power and Spin Observable}

The analyzing power $A_y$ and the spin observable (Q-Value) are calculated 
from the general formulae given in Eqns. (41) and (42) respectively. The 
results of our calculations for $\overrightarrow{p}+ ^{40}Ca$ system at 
incident energies 300 MeV and 800 MeV are shown in Figs. 8 and 9. The RIA 
predictions for $A_y$ using RLF with RMF (NL3) density show a 
quantitative agreement with the data \cite{coop93} at 300 MeV whereas this 
observation is just reverse at 800 MeV \cite{ferg86}. That means, the 
prediction of $A_y$ resemble the $\frac{d\sigma}{d\Omega}$ observations of Figs. 
$5-7$. In Figs. 10 and 11, we present the $A_y$ and $Q-$value for 
$\overrightarrow{p}+ ^{40,42,44,48}Ca$ composite system at 1000 MeV. 
These results are obtained for both the  RLF and MRW parametrizations with NL3 and 
G2 densities in comparision with the experimental data \cite{expt78}. The calculated 
$A_y$ and $Q-$values obtained by these two forces differ significantly from each 
other for the choice of RLF and MRW parametrizations. Also, we observe small oscillations
in the values of $A_y$ and $Q$ with 
the increase in scattering angle $\theta_{c.m.}$ for both RLF and MRW. This oscillatory 
behavior could be related with the dispersion phenomenon of the optical potential. 
Similar to the $\frac{d\sigma}{d\Omega}$, here also the prediction of MRW is best fitted 
to the data for the higher and RLF for lower incident energies. Further, investigation 
into the spin rotation parameter $Q-$value, the peak shift and diminished magnitude 
with the increase in neutron number (see Fig. 11) agrees with the calculation of first order 
Brueckner theory using Urbana V14 soft core inter-nucleon interactions \cite{brue09}. 
It leads to the nucleon finite size correction more realistic and hence merits a 
structure effect for the formation of exotic nuclei in laboratory.

\section{Summary and Conclusion}

We have calculated the density distribution of protons and neutrons 
for $^{40,42,44,48}Ca$ by using RMF (NL3) and E-RMF (G2) parameter sets. 
From these densities, we estimate the relative isotopic neutron density
difference for both the force parameters. The comparison of $\Delta\rho_n(r)$ 
with the data \cite{ray81} indicates the superiority of G2 over NL3.
The small difference in the density at the central region significantly affect
the results of scattering observables including the optical potential.
A fundamental difference between RLF and MRW parametrizations as well as RMF (NL3)
and E-RMF (G2) sets in the RIA predictions is noticed from the observation of 
$\frac{d\sigma}{d\Omega}$, $A_y$ and Q-value. We conclude from our calculations that
RLF relatively works well at lower and MRW at higher incident energies. The predicting
capability of scattering observables of RMF (NL3) over E-RMF (G2) is also realised. 

In conclusion, the reaction dynamics highly depends on the input density and 
the choice of parametrization. In addition to this, our present study indicates that
the RIA is a powerful predictive model which provides a clear picture about the successful 
Dirac optical potentials and can be useful to study both stable and exotic nuclei.

\section*{Acknowledgments}

We thank Dr. BirBikram Singh for a careful reading of the manuscript. 
This work is supported in part by UGC-DAE Consortium for Scientific Research,
Kolkata Center, Kolkata, India (Project No. UGC-DAE CRS/KC/CRS/2009/NP06/1354).

\bigskip

\appendix
\section{}

If RLF is our choice, the ${\it t}^L$ functions in Eqns. (17-19) and 
(23-24) involves all the occupied states for ${\it pp}$ and ${\it pn}$ 
scattering. It is most convenient to shift variables from 
$x'\rightarrow x+x'$ so the  ${\it t}^L$ functions are not involved in 
the angular integration. Now, the first order optical potential Eqn. (20) can 
be written as \cite{horo90},

\begin{widetext}
\begin{eqnarray}
U^{L}(r, E) & = & \frac{-4\pi ip}{M}\left[ \int d^{3}r^{'} \rho^{L}(x+x') t^{L}_{D}
(r'; E)\right] \nonumber\\
&& + \left[ \int d^{3}r' \rho^{L}(x+x', x) 
t^{L}_{X}(r'; E)j_{0}(pr')\right],
\end{eqnarray}
after $\phi$ integration, this become
\begin{eqnarray}
U^{L}(r, E) & = & \frac{-8\pi^2 ip}{M}\left[ \int dr't^{L}_{D}(r'; E)
\int^{+1}_{-1}\rho^{L}(x+x')d\omega \right] \nonumber\\
&& + \left[ \int dr' t^{L}_{X}(r'; E)j_{0}(pr')
\int^{+1}_{-1}\rho^{L}(x+x', x)d\omega\right] ,
\end{eqnarray}
\end{widetext}
where $\omega=cos\theta$, $(\vert x+x'\vert^2)=(r^2+r'^2+2\omega r r')$ and 
$(\frac{\vert {2x+x'}\vert}{\vert 4 \vert}^2)=\frac{1}{4} (r'^2+4\omega r r' +4r^2)$. 
The integral evaluated by Gauss-Laguerre quadrature. At the point $(x+x')$, the 
radial integration must go roughly twice the nuclear radius. Note that 
for spherical nuclei only the scalar and vector are taken into account, 
as the tensor terms are negligible.

In case of MRW, the optical potential $U_{opt}$ is calculated somewhat differently 
from the RLF. Here we tranform the density $\rho^L(x)$ to momentum space, then multiply 
with the ${\cal F}^L(q,E)$, and back which leads to the equation

\begin{widetext}
\begin{eqnarray}
U^{L}(r, E) = \frac{-4\pi ip}{M}\left[ \int \frac {d^{3}q}{(2\pi)^3} e^{iqx} 
{\cal F}^L(q,E) \int d^3x' e^{-iqx'}\rho^{L}(r')\right ],
\end{eqnarray}
with ${\cal F}^L(q,E)={\cal F}^L_0(E) e^{-q^2\beta^2(E)}$ at each proton energy E. 
The final equation is obtained by adding the contributions from proton and neutron 
states to the direct term Eqn. (A3) which is given as,

\begin{eqnarray}
U^{L}(r, E) = \frac{-8 ip}{M} \frac{1}{r}\left[ \int_0^\infty dq Sin (qr) 
\int_0^\infty dr' r' Sin (qr') {\cal F}^L_0(E) e^{-q^2\beta^2(E)} \rho^{L}(r')\right ].
\end{eqnarray}
\end{widetext}

The above integrals is solved by double Gussian summation methods.

\end{document}